\documentclass[11pt]{article}
\usepackage{amsmath,amssymb,amsthm,mathrsfs,stackrel}
\usepackage{array}
\usepackage{graphicx,psfrag,epsf}
\usepackage{multirow}
\usepackage{multicol}
\usepackage{threeparttable}
\usepackage{booktabs}
\usepackage[round]{natbib}
\usepackage{float}
\usepackage{algorithmic}
\usepackage[ruled,vlined]{algorithm2e}
\usepackage{url}
\setcitestyle{authoryear,round}

\makeatletter
\renewcommand{\algocf@captiontext}[2]{#1\algocf@typo. \AlCapFnt{}#2}

\def\@algocf@capt@plain{top}
\renewcommand{\algocf@makecaption}[2]{%
  \addtolength{\hsize}{\algomargin}%
  \sbox\@tempboxa{\algocf@captiontext{#1}{#2}}%
  \ifdim\wd\@tempboxa >\hsize%
    \hskip .5\algomargin%
    \parbox[t]{\hsize}{\algocf@captiontext{#1}{#2}}%
  \else%
    \global\@minipagefalse%
    \hbox to\hsize{\box\@tempboxa}%
  \fi%
  \addtolength{\hsize}{-\algomargin}%
}
\makeatother

\usepackage{amsfonts,bm,amsbsy,bbm}
\usepackage{rotating,lscape}
\usepackage{booktabs}
\usepackage{xcolor}
\usepackage{xr-hyper}
\usepackage[colorlinks,citecolor=blue,linkcolor=blue,urlcolor=blue]{hyperref}

\def\X{{\boldsymbol X}}
\def\Z{{\boldsymbol Z}}
\def\U{{\boldsymbol U}}
\def\V{{\boldsymbol V}}

\def\x{{\boldsymbol x}}
\def\z{{\boldsymbol z}}
\def\J{{\boldsymbol J}}
\def\c{{\boldsymbol c}}
\def\m{{\boldsymbol m}}
\def\bmu{{\boldsymbol \mu}}
\def\bbeta{{\boldsymbol \beta}}
\def\beeta{{\boldsymbol \eta}}
\def\btheta{{\boldsymbol \theta}}
\def\bgamma {{\boldsymbol \gamma}}
\def\bSigma{{\boldsymbol \Sigma}}

\def\bxi{{\boldsymbol \xi}}
\def\zero{{\boldsymbol 0}}

\def\a{{\boldsymbol a}}

\def\bvartheta{{\boldsymbol \vartheta}}

\def\bphi{{\boldsymbol \phi}}
\def\bpsi{{\boldsymbol \psi}}

\newtheorem{assumption}{Assumption}

\newtheorem{theorem}{Theorem}
\newtheorem{proposition}{Proposition}

\newcommand{\w}{\mathbbm{w}}
\newcommand{\E}{\mathbb{E}}
\newcommand{\R}{\mathbb{R}}
\def\t{\top}
\def\s{\sigma}
\def\eps{\epsilon}
\def\l{\left}
\def\r{\right}
\def\fone{\frac{1}{n}}
\def\fNone{\frac{1}{N}}
\def\sumn{\sum^n_{i=1}}
\def\sumN{\sum^{n+N}_{i=n+1}}

\begin{document}

\title{Augmented transfer regression learning for completely missing covariates}

\author{
Huali Zhao\textsuperscript{1} and 
Tianying Wang\textsuperscript{2}\thanks{Tianying.Wang@colostate.edu}\\[0.5em]
\textsuperscript{1}School of Mathematics and Statistics, Huazhong University of Science and Technology\\
\textsuperscript{2}Department of Statistics, Colorado State University
}

\date{}

\maketitle
\thispagestyle{empty}
\baselineskip=20pt

\begin{abstract}
Large-scale population-level datasets, such as the UK Biobank and the All of Us Research Program, often lack covariates needed for a specific analysis, such as genetic or lifestyle measures, while related studies measure them. This creates a cross-population missing data problem in which covariates are completely unobserved in the target population, rather than partially missing within one dataset. We propose an augmented transfer regression learning method for this setting. The key identifying condition is a sub-population shift assumption: the joint distribution of the outcome and observed covariates may differ across source and target populations, but the conditional distribution of the missing covariates given observed variables is invariant. We combine importance-weighted estimating equations with imputation terms for first- and second-order moments of the missing covariates. The resulting estimator is doubly robust and remains consistent if either the density ratio model or both imputation models are correctly specified. It is \(n^{1/2}\)-consistent and asymptotically normal, and attains the semiparametric efficiency bound when both nuisance models are correctly specified. 
\end{abstract}


\pagestyle{plain}

\begingroup
\allowdisplaybreaks
\section{Introduction}
Modern biomedical research increasingly uses large-scale population resources, such as the UK Biobank \citep{conroy2023uk} and the All of Us Research Program \citep{denny2019all}. These general-purpose resources are not designed for every downstream analysis, and variables needed for a particular scientific question may be absent. We consider the setting in which some covariates of interest are completely unobserved in the target population but available in a related source population. This differs from conventional missing data settings, where covariates are at least partially observed in the analysis population. 

For example, health surveys such as the Behavioral Risk Factor Surveillance System \citep{mokdad2009behavioral} and the National Health and Nutrition Examination Survey \citep[NHANES]{johnson2013national} collect behavioural variables and outcomes, but typically do not include genetic measures such as polygenic risk scores. Conversely, genetic resources may not contain the same detailed behavioural or environmental variables as specialised health surveys. These examples raise the statistical question of how to use source-population information to estimate target-population regression parameters while preserving valid inference under possible source-target distributional shifts.

Classical missing data methods focus on estimation when missingness occurs within a single analysis population, under missing completely at random, missing at random, or missing not at random mechanisms \citep{rubin1976inference,yang2019causal}. Estimating equation methods under ignorable missingness \citep{zhao1996regression,lipsitz1999weighted}, as well as doubly robust and multiply robust estimators \citep{bang2005doubly,han2014multiply}, exploit partial observability of the covariates within the analysis population. More recently, \citet{kluger2025prediction} proposed a predict-then-debias method for missing covariates, using a small complete sample from the target population to correct bias from imputation. These approaches are not directly applicable when some covariates are completely unmeasured in the target population and only available from a distinct source population.

Transfer learning provides another natural perspective, but most existing methods address missing outcomes or labels rather than missing covariates. Under covariate shift, doubly robust estimating equations typically assume invariant conditional outcomes across populations and estimate a density ratio using covariates observed in both populations \citep{sugiyama2008direct,kpotufe2021marginal,liu2023augmented,zhou2024domain,zhou2025doubly}. Under label shift, doubly flexible methods assume invariant conditional covariate distributions given the outcome \citep{lipton2018detecting,garg2020unified,lee2025doubly}. In our setting, however, some covariates needed in the target estimating equation are entirely missing in the target population. Hence, a density ratio involving the full covariate vector is not estimable from the target data, and the target estimating equation requires first- and second-order conditional moments of the missing covariates. 
Beyond these strands, general semiparametric theory for moment models with auxiliary data \citep{chen2008semiparametric,yan2024transfer} accommodates variables that are unavailable in one sample and establishes the corresponding efficiency bounds; the resulting constructions, however, treat the conditional moment of the estimating function as a generic, parameter-dependent nuisance and do not exploit the regression structure of the problem considered here. Together, these features define a cross-population completely missing covariate problem that none of the above approaches addresses directly.

We propose an augmented transfer regression learning method to address completely missing covariates in the target population. Our main contributions are summarised as follows:
\begin{enumerate}
    \item We identify and formulate a cross-population completely missing covariate problem. In this setting, the covariates of interest are absent for all target observations but are observed, together with the outcome and other covariates, in a related source population, and the source and target distributions may differ. This problem is distinct from classical missing data settings, which rely on partial observability within the analysis population, and from transfer learning methods for missing outcomes. We construct augmented estimating equations that combine importance weighting with imputation terms for the first- and second-order moments of the missing covariates, yielding a doubly robust estimator of the target-population regression parameter.
    
    \item We introduce a sub-population shift assumption, under which the joint distribution of the outcome \(Y\) and covariates \(\Z\) observed in both populations may differ between the source and target populations, while the conditional distribution of the target-missing covariates \(\X\) given \((Y,\Z)\) is invariant: \(p_0(\x\mid y,\z)=p_1(\x\mid y,\z)\), where \(0\) and \(1\) index the target and source populations. This condition provides the source-target link needed to identify the target estimating equation when \(\X\) is completely missing in the target population. From a missing-data perspective, this condition has the missing at random (MAR)-type form \(\X\perp S\mid(Y,\Z)\). In contrast to the usual MAR setting, where $\X$ is observed for a subset of the analysis population, $\X$ here is completely absent from the target sample, so that \(p(\x\mid y,\z)\) must be learned from a separate source population.
    
   \item We establish \(n^{1/2}\)-consistency and asymptotic normality under double robustness: the estimator remains consistent if either the density ratio model is correctly specified or both imputation models are correctly specified. When both nuisance models are correctly specified, we further show that the estimator attains the semiparametric efficiency bound.
\end{enumerate}

The rest of the paper is organised as follows. Section~\ref{sec:method} introduces the proposed estimating equations and nuisance-model estimation. Section~\ref{sec:theory} establishes consistency, asymptotic normality, and semiparametric efficiency. Section~\ref{sec:simulation} presents simulation studies, and Section~\ref{sec:data analysis} applies the method to UK Biobank data. Section~\ref{sec:discussion} concludes with discussion and extensions.

\section{Methodology}\label{sec:method}
\subsection{Problem statement}
Let \(S=0\) denote the target population. A target observation contains the response \(Y\) and covariates \(\Z=(Z_1,\ldots,Z_q)^\top\), whereas the covariates of interest \(\X=(X_1,\ldots,X_p)^\top\) are unobserved. Let \(S=1\) denote the source population, where \((Y,\X,\Z)\) is observed. We first propose the \textbf{sub-population shift assumption}:
\begin{equation}\label{assum:joint pdf}
    (Y,\X,\Z) \mid S=s \sim p_s(y,\z)p(\x \mid y,\z),
\end{equation}
Here, \(p_s(y, \z)\) and \(p(\x \mid y, \z)\) denote, with respect to suitable dominating measures, the joint and conditional densities or probability mass functions of \((Y, \Z) \mid S = s\) and \(\X\mid(Y,\Z)\), respectively; the latter is assumed to be invariant across the two populations. For example, the target NHANES data include smoking status, alcohol consumption, physical activity, dietary intake ($\Z$), and body mass index (BMI, $Y$), but lack polygenic risk scores for BMI ($X$). In contrast, the UK Biobank contains all these variables ($X,Y,\Z$). Because $X$ is absent from NHANES by design, there is no subject-level selection into its measurement. By contrast, the equality of $p(x\mid y,\z)$ between NHANES and UK Biobank remains a separate cross-population transportability assumption. Its plausibility depends on whether population-specific factors related to $X$ are adequately captured by $(Y,\Z)$.

From the transfer learning perspective, assumption~\eqref{assum:joint pdf} is more flexible and less restrictive than the commonly used label shift assumption in the context of distribution shift with completely missing outcomes \citep{lee2025doubly}. Specifically, label shift assumes $(Y, \X, \Z) \mid S = s \sim p_s(y) p(\x, \z\mid y)$, which implies that both the conditional density of $(\X,\Z)$ given $Y$ and the conditional density of $\Z$ given $Y$ remain invariant between the two populations. Thus, the sub-population shift assumption holds as $p(\x\mid y,\z)=p(\x,\z\mid y)/p(\z\mid y)$. Additionally, assumption~\eqref{assum:joint pdf} can be viewed from the missing data perspective. By treating $\X$ as observed when $S=1$ and missing when $S=0$, it corresponds to the MAR condition $\X\perp S\mid Y,\Z$. We note that MAR itself allows the distribution of the observed variables $(Y,\Z)$ to differ across missingness patterns. Thus, assumption~\eqref{assum:joint pdf} is more appropriately viewed as an MAR-type conditional transportability condition in a two-sample setting, where $p(\x\mid y,\z)$ is transported from the source to the target while $p_s(y,\z)$ may differ across populations. This dual interpretation connects assumption~\eqref{assum:joint pdf} to both transfer learning and missing data literature.

Our primary interest lies in estimation and inference for the coefficient vector in the target-population working model: 
\begin{equation}\label{eq:workingmodel}
    \E_0(Y \mid \X,\Z) = \X^\t\bbeta + \Z^\t\btheta,
\end{equation}
where $\E_s$ is the expectation operator on the population $S=s$ for $s=0,1$, and $\bvartheta:=(\bbeta^\t,\btheta^\t)^\t\in\R^{p+q}$ is the coefficient vector of interest. Without loss of generality, we assume that the intercept term is included in \(\btheta\).

Define the parameter $\bvartheta_0:=(\bbeta_0^\t,\btheta_0^\t)^\t\in\R^{p+q}$ as the solution to the estimating equation in the target population $S=0$:
\begin{equation}\label{eq:estimating_eq}
    \E_0\big\{(\X^\t,\Z^\t)^\t (Y - \X^\t\bbeta - \Z^\t\btheta)\big\} = \zero.
\end{equation}
Because of the sub-population shift, the joint distribution of $(Y,\Z)$ in the source data may differ from that in the target data. Hence, even if the working model~\eqref{eq:workingmodel} holds in the target data, $\E_1\{(\X^\t,\Z^\t)^\t (Y - \X^\t\bbeta - \Z^\t\btheta)\}$ need not be zero in the source population. Along with potential model misspecification of the working model~\eqref{eq:workingmodel}, directly solving the empirical estimating equation in Eq~\eqref{eq:estimating_eq} using source data to estimate $\bvartheta_0$ often yields inconsistent estimators.

\subsection{Two preliminary methods}
To motivate our doubly robust estimator, we present two preliminary methods in this section: the importance weighting method and the imputation method. Based on our sub-population shift assumption~\eqref{assum:joint pdf}, an intuitive method for estimating $\bvartheta_0$ is to multiply the source estimating function in Eq~\eqref{eq:estimating_eq} by the importance weight $p_0(Y,\Z)/p_1(Y,\Z)$ and solve its empirical counterpart. Another intuitive method is to impute the absent $\X$ in the target data and then plug it into the estimating equation~\eqref{eq:estimating_eq}. However, imputing the first- and second-order moments of $\X$ in Eq~\eqref{eq:estimating_eq} is more challenging than in the context of completely missing outcomes, which only requires imputing the first moment of $Y$ \citep{liu2023augmented,zhou2025doubly}. 

The first preliminary method is to incorporate importance sampling weights and solve the weighted estimating equations using the source data ($S=1$),
\begin{equation}\label{eq:iw_estimating}
    \E_1\big\{\omega(Y,\Z)(\X^\t,\Z^\t)^\t(Y-\X^\t\bbeta-\Z^\t\btheta)\big\} = \zero,
\end{equation}
where $\omega(Y,\Z)$ is a working model for the density ratio $\w(Y,\Z) := p_0(Y,\Z)/p_1(Y,\Z) \in \R_{+}$. We assume the usual overlap condition, namely \(\w(Y,\Z)\le C<\infty\) almost surely for some constant \(C\).

Assume the target data ($S=0$) consist of $N$ independent and identically distributed observations on $Y$ and $\Z$ only, while the source data ($S=1$) consist of $n$ independent and identically distributed observations on $(Y,\X,\Z)$. Write the full observed sample as $\{(Y_i,S_i\X_i,\Z_i,S_i):i=1,\ldots,n+N\}$. We index the source observations first, so $S_i=I(i\le n)$, where $I(\cdot)$ denotes the indicator function; the remaining $N$ observations are from the target population. We have the following proposition for obtaining a consistent estimator of $\bvartheta_0$.
\begin{proposition}\label{prop:IW}
    If $\omega(Y,\Z)$ is correctly specified for the density ratio $\w(Y,\Z)$ and $\widehat\bvartheta_{\rm IW}:=(\widehat\bbeta_{\rm IW}^\t,\widehat\btheta_{\rm IW}^\t)^\t$ denotes a solution to the empirical counterpart of Eq~\eqref{eq:iw_estimating}, then the importance weighting {\rm (IW)} estimator $\widehat\bvartheta_{\rm IW}$ is consistent for $\bvartheta_0$.
\end{proposition}

The second preliminary method is based on imputation models. Let $\m_1(Y,\Z):\ \R \times \R^q \to \R^p$ and $\m_2(Y,\Z):\ \R \times \R^q \to \R^{p \times p}$ denote the imputation models used respectively to approximate the following two models:
\begin{align*}
    \bmu_1(y,\z) &= \E(\X \mid Y=y, \Z=\z) = \E_s(\X \mid Y=y, \Z=\z),\\
    \bmu_2(y,\z) &= \E(\X\X^\t \mid Y=y, \Z=\z) = \E_s(\X\X^\t \mid Y=y, \Z=\z),
\end{align*}
where $s=0,1$. It is natural to estimate $\bvartheta_0$ by solving the following estimating equations: 
\begin{equation}\label{eq:imp_estimating}
    \E_0\big\{\m_1(Y,\Z)(Y - \Z^\t\btheta)- \m_2(Y,\Z)\bbeta\big\} = \zero,\ \E_0\big\{\Z(Y - \m_1(Y,\Z)^\t\bbeta - \Z^\t\btheta)\big\} = \zero,
\end{equation}
which leads to the following proposition for an alternative consistent estimator.
\begin{proposition}\label{prop:IMP}
    If $\m_{\iota}(Y,\Z)$ is correctly specified for $\bmu_{\iota}(Y,\Z)$, $\iota=1,2$, and $\widehat\bvartheta_{\rm IMP}:=(\widehat\bbeta_{\rm IMP}^\t,\widehat\btheta_{\rm IMP}^\t)^\t$ denotes a solution to the empirical counterpart of Eq~\eqref{eq:imp_estimating}, then the imputation {\rm (IMP)} estimator $\widehat\bvartheta_{\rm IMP}$ is consistent for $\bvartheta_0$.
\end{proposition}

However, the IMP estimator $\widehat\bvartheta_{\rm IMP}$ relies on the model specification of \(\m_{\iota}(Y,\Z)\) for \(\bmu_{\iota}(Y,\Z), \iota=1,2\). Similarly, the validity of the IW estimator $\widehat\bvartheta_{\rm IW}$ heavily depends on the model specification of \(\omega(Y,\Z)\) for \(\w(Y,\Z)\). Our numerical experiments indicate that both preliminary methods perform poorly when the nuisance models are misspecified or poorly estimated (Table~\ref{tab:point-var} and Table~\ref{tab:dataresults}).

\subsection{The proposed doubly robust method}
We now propose an augmented transfer regression learning method to overcome the limitations of using only the IW or IMP methods. The key is to combine importance weighting with imputation so the resulting estimating equations do not rely fully on either nuisance model alone. We use the working conditional-moment models $\m_1(Y,\Z),\m_2(Y,\Z)$, imputing the first- and second-order moments of the missing $\X$ for the target data, and augment the importance sampling weighted estimating equation~\eqref{eq:iw_estimating} with the imputed data, which results in the following augmented estimating equations,
\begin{align}
    \U_{\rm DR}(\bvartheta) 
    & \equiv \E_1\left(\omega(Y,\Z)\left[\{\X - \m_1(Y,\Z)\}(Y - \Z^\t\btheta) + \{\m_2(Y,\Z)-\X\X^\t\}\bbeta\right]\right) \notag \\
    & \quad + \E_0\{\m_1(Y,\Z)(Y-\Z^\t\btheta) - \m_2(Y,\Z)\bbeta\} = \zero,\label{eq:DR_estimating_1}\\
    \V_{\rm DR}(\bvartheta) 
    & \equiv \E_1[\omega(Y,\Z)\Z\{\m_1(Y,\Z) - \X\}^\t\bbeta] + \E_0[\Z\{Y - \m_1(Y,\Z)^\t\bbeta - \Z^\t\btheta\}] = \zero.\label{eq:DR_estimating_2}
\end{align}

The construction of Eqs~\eqref{eq:DR_estimating_1}-\eqref{eq:DR_estimating_2} aligns with the existing literature on doubly robust estimators for the average treatment effect in causal inference studies \citep[Chapter 12]{ding2024first}, as well as doubly robust estimators in transfer learning settings with missing outcomes or labels in the target data \citep{liu2023augmented,zhou2024domain,zhou2025doubly}. It is important to emphasise that the scenario of completely missing covariates is more challenging and intricate than that of missing outcomes, due to the distinct roles covariates and outcomes play in estimating equations. Unlike the single imputation model used for missing $Y$ in \citet{liu2023augmented,zhou2024domain,zhou2025doubly}, our approach employs two imputation models, \(\m_{\iota}, \iota=1,2\), to estimate the first- and second-order moments of the missing \(\X\) in the target data, which adds complexity and makes it challenging to ensure the double robustness properties of the estimating equation. 

\begin{theorem}[Double robustness]\label{thm:DR}
     The augmented estimating equations~\eqref{eq:DR_estimating_1}-\eqref{eq:DR_estimating_2} are doubly robust to the specification of nuisance models: $\omega(Y,\Z)$ and $\m_\iota(Y,\Z), \iota=1,2$.
\end{theorem}
Theorem~\ref{thm:DR} motivates estimating \(\bvartheta_0\) by solving empirical counterparts of the augmented estimating equations~\eqref{eq:DR_estimating_1}-\eqref{eq:DR_estimating_2}. Let
\(\widehat\bvartheta_{\rm DR}:=(\widehat\bbeta_{\rm DR}^\top,\widehat\btheta_{\rm DR}^\top)^\top\)
denote the resulting estimator. Then \(\widehat\bvartheta_{\rm DR}\) is doubly robust for
\(\bvartheta_0=(\bbeta_0^\top,\btheta_0^\top)^\top\): it is consistent if either the density ratio model \(\omega(Y,\Z)\) is correctly specified or the imputation models \(\m_1(Y,\Z)\) and \(\m_2(Y,\Z)\) are correctly specified.
For variance estimation, we use a stratified nonparametric bootstrap that resamples the source and target observations separately, preserving the sample sizes \(n\) and \(N\). In each bootstrap replication, we re-estimate the density ratio model \(\widehat\omega\) using the resampled \((Y,\Z)\) values from both populations, re-estimate the imputation models \(\widehat\m_1\) and \(\widehat\m_2\) using the resampled source observations, and then solve empirical versions of Eqs~\eqref{eq:DR_estimating_1}-\eqref{eq:DR_estimating_2}. The bootstrap covariance estimator is the empirical covariance of the resulting bootstrap estimates.
Although Theorem~\ref{thm:main} provides an asymptotic variance expression, direct plug-in estimation requires estimating multiple covariance terms, derivatives of the estimating equations, and nuisance-model quantities such as \(\m_1\), \(\m_2\), and \(\omega\). We use the bootstrap because it accounts for nuisance re-estimation directly and is often more stable in finite samples for related transfer regression problems \citep{liu2023augmented, li2024adaptive, zhou2025doubly}.

\subsection{Estimation of nuisance models}\label{subsec:nuisance} 
We now describe how we estimate the nuisance models: $\omega(Y,\Z)$ and $\m_{\iota}(Y,\Z),\iota=1,2$. 
To adjust for source-target differences in the distribution of $(Y,\Z)$, we posit the following working density-ratio model for $\w(Y,\Z)$:
\begin{align}\label{eq:densityratio}
    \omega(y,\z) = \exp(y\eta_y+\z^\t\beeta_z),
\end{align}
where $\eta_y\in\R$ and $\beeta_z\in\R^q$ are nuisance parameters. If $\omega(y,\z)$ is correctly specified, $\E_1\{\omega(Y,\Z)(Y,\Z^\t)^\t\} = \E_0\{(Y,\Z^\t)^\t\}$. Thus, we define the population parameters $\bar\beeta:=(\bar\eta_y,\bar\beeta_z^\t)^\t$ as 
$
   \bar\beeta = \arg\min_{\eta_y\in\R,\beeta_z\in\R^q} \E_1\{\exp(Y\eta_y+\Z^\t\beeta_z)\} - \E_0(Y\eta_y+\Z^\t\beeta_z).
$
Since $\bar\beeta$ can be estimated with data $(Y,\Z)$, which is often of large size and observable across two populations, a consistent estimator for $\bar\beeta$ can be constructed as
\[
     \widehat\beeta:=(\widehat\eta_y,\widehat\beeta_z^\t)^\t = \arg\min_{\eta_y\in\R,\beeta_z\in\R^q} \l\{\fone\sum^n_{i=1} \exp(Y_i\eta_y+\Z_i^\t\beeta_z) - \fNone\sum^{n+N}_{i=n+1} (Y_i\eta_y+\Z_i^\t\beeta_z)\r\}.
\]
Thus, the estimated density ratio is $\widehat\omega(y,\z)=\exp(y\widehat\eta_y+\z^\t\widehat\beeta_z)$.

To estimate the imputation models $\m_\iota(Y,\Z)$, $\iota=1,2$, we use a working conditional density $\mu(\X\mid Y,\Z;\widehat\bgamma)$ as a convenient device for constructing the first two conditional moments, where $\mu$ is a pre-specified parametric model and $\widehat\bgamma$ represents empirical estimates of the unknown model parameters. Correct specification of the full conditional density is not required; it is sufficient that the induced first two conditional moments are correctly specified. Specifically, the imputation models are constructed as:
\[
    \m_1(Y,\Z;\widehat\bgamma) = \int \x \mu(\x\mid Y,\Z; \widehat\bgamma)d\x,\quad \m_2(Y,\Z;\widehat\bgamma) = \int \x\x^\t \mu(\x\mid Y,\Z; \widehat\bgamma)d\x.
\]

We present an illustrative example to explain the construction strategy and practical computation for \( \widehat{\mu}(\x\mid y,\z) := \mu(\x \mid y, \z; \widehat{\bgamma}) \) and \( \widehat\m_{\iota}(Y, \Z) := \m_{\iota}(Y, \Z; \widehat{\bgamma}), \, \iota = 1, 2 \). Specifically, we consider the following data-generating model
\[
    \mu(\x\mid y,\z;\bgamma) = \frac{1}{(2\pi)^{p/2}|\bSigma_{x,yz}|^{1/2}}\exp\l\{-\frac{1}{2}(\x - \bgamma_yy-\bgamma_z\z)^\t\bSigma_{x,yz}^{-1}(\x - \bgamma_yy-\bgamma_z\z)\r\},
\]
where $\bgamma:=(\bgamma_y,\bgamma_z,\bSigma_{x,yz})\in\R^{p\times (1+q+p)}$ is the parameter matrix. In other words, we posit $\X\mid Y,\Z\sim N_p(\bgamma_yY+\bgamma_z\Z,\bSigma_{x,yz})$. We estimate $\bgamma$ by maximum likelihood and write the resulting estimator as $\widehat\bgamma:=(\widehat\bgamma_y,\widehat\bgamma_z,\widehat\bSigma_{x,yz})$. Subsequently, $\widehat\m_1(Y,\Z) = \widehat\bgamma_yY+\widehat\bgamma_z\Z$ and $\widehat\m_2(Y,\Z) = \widehat\bSigma_{x,yz} + \widehat\m_1(Y,\Z)\widehat\m_1(Y,\Z)^\t$. More generally, $\mu(\x\mid y,\z;\bgamma)$ may be specified from an elliptically contoured family, with the multivariate normal as a special case. 

We also discuss alternative working models for the density ratio and the imputation functions, and we provide theoretical guarantees when these nuisance functions are estimated with flexible machine learning methods; see Appendices~A2-A3 for details.

\subsection{Comparison with existing transfer learning methods for completely missing outcomes}
In the transfer learning literature developed for completely missing outcomes ($Y$), \citet{liu2023augmented} considered the covariate shift assumption, i.e., \( p_0(y \mid \x) = p_1(y \mid \x) \) and \( p_0(\x) \neq p_1(\x) \). Then, the density ratio is defined as \( \w(\x) = p_0(\x) / p_1(\x) \) and the outcome imputation model is \( \m(\x) = \E(Y \mid \X = \x) = \E_s(Y \mid \X = \x) \) where $s=0,1$. Their doubly robust estimating equation is expressed as: $\E_1[\w(\X)\X\{Y-\m(\X)\}] + \E_0[\X\{\m(\X)-\X^\t\bbeta\}] = \zero$. In this missing-outcome setting, \( \w(\X) \) is estimable based on \emph{observed \(\X\) in both populations}. Since only one imputation model is needed to impute $Y$, that is, $\m(\X)$, their estimating equation remains concise and can be computationally straightforward. Related transfer-learning methods for missing outcomes have also been
developed by \citet{zhou2025doubly}.

Table~\ref{tab:comparison_transfer} compares our method with
\citet{liu2023augmented}, whose formulation provides a close structural analogue in the completely missing-outcome setting, in terms of identification, nuisance functions, estimating equations, and semiparametric efficiency.
First, because \(\X\) is not observed in the target population, the usual covariate-shift density ratio \(\w(\X)\) is not estimable from the available data. Our sub-population shift assumption instead leads to the estimable density ratio \(\w(Y,\Z)\). Second, rather than using a single conditional mean model for a missing outcome, our method requires models for both \(\E(\X\mid Y,\Z)\) and \(\E(\X\X^\t\mid Y,\Z)\). The coefficients of \(\X\) and \(\Z\) are consequently estimated through two coupled estimating equations, which also leads to a more involved asymptotic analysis. When both nuisance models are correctly specified, we further establish semiparametric efficiency. 
\citet{chen2008semiparametric} derived efficiency bounds and efficient estimators for parameters defined by general moment restrictions in an auxiliary-data framework, including a verify-out-of-sample setting in which the variables of interest are completely unavailable in the primary population, while \citet{yan2024transfer} established semiparametric efficiency for general estimating equations under covariate-shift transfer learning. Our efficiency result instead characterises the bound for the completely missing-covariate problem under the sub-population shift assumption.

\begin{table}[!ht]
\caption{Comparison with a closely related transfer learning method
for completely missing outcomes.}
\label{tab:comparison_transfer}
\centering
\small
\begin{tabular}{
    >{\raggedright\arraybackslash}m{3.0cm}
    >{\raggedright\arraybackslash}m{5.0cm}
    >{\raggedright\arraybackslash}m{5.0cm}
}
\toprule
&
\shortstack[l]{
    \textbf{Missing outcomes}\\[0.4ex]
    {\footnotesize \cite{liu2023augmented}}
}
&
\shortstack[l]{
    \textbf{Missing covariates}\\[0.4ex]
    Proposed method
}
\\
\midrule

Identifying assumption
& $p_0(y\mid\x)=p_1(y\mid\x)$
& $p_0(\x\mid y,\z)=p_1(\x\mid y,\z)$ \\[1.3ex]

Density ratio
& $\w(\x)=p_0(\x)/p_1(\x)$
& $\w(y,\z)=p_0(y,\z)/p_1(y,\z)$ \\[1.3ex]

Imputation
& $\E(Y\mid\X)$
& $\E(\X\mid Y,\Z),\ \E(\X\X^\t\mid Y,\Z)$ \\[1.3ex]

Estimating equations
& One estimating equation
& Two coupled estimating equations \\[1.3ex]

Semiparametric efficiency
& Not established
& Established \\
\bottomrule
\end{tabular}
\end{table}

\section{Theoretical properties}\label{sec:theory}
For any vector $\a$, let $\|\a\|_2$ represent its $\ell_2$ norm and define $\a^{\otimes 2}:=\a\a^\t$. Assume that the dimensionalities of $\X$ and $\Z$, that is, $p$ and $q$, are fixed.
We treat the source and target sample sizes \(n\) and \(N\) as fixed by design and consider an asymptotic sequence in which \(n,N\to\infty\) and \(n/(n+N)\to\rho\in(0,1)\).
Let $\bar\bgamma$ denote the probability limit of $\widehat\bgamma$. Denote $\bar\omega(Y,\Z)=\exp(Y\bar\eta_y+\Z^\t\bar\beeta_z)$ and $\bar\m_{\iota}(Y,\Z):=\m_{\iota}(Y,\Z;\bar\bgamma)$, $\iota=1,2$. Let $\U(\bvartheta)$ and $\V(\bvartheta)$ denote the population estimating functions in Eqs~\eqref{eq:DR_estimating_1}-\eqref{eq:DR_estimating_2} with the nuisance functions fixed at $\bar\omega$, $\bar\m_1$, and $\bar\m_2$.
Define the information matrix
\[
\small
\J_{\bvartheta}:= -
    \begin{pmatrix}
        \E\{\frac{\partial \U(\bvartheta)}{\partial\bbeta}\} & \E\{\frac{\partial \U(\bvartheta)}{\partial\btheta}\} \\
        \E\{\frac{\partial \V(\bvartheta)}{\partial\bbeta}\} & \E\{\frac{\partial \V(\bvartheta)}{\partial\btheta}\} 
    \end{pmatrix}
    \in\R^{(p+q)\times (p+q)},
\]
where 
\begin{align*}
\small
    \E\l\{\frac{\partial \U(\bvartheta)}{\partial\bbeta}\r\} 
    & = \E_1[\bar\omega(Y,\Z)\{\bar\m_2(Y,\Z) - \X\X^\t\}] - \E_0\{\bar\m_2(Y,\Z)\},\\
    \E\l\{\frac{\partial \V(\bvartheta)}{\partial\bbeta}\r\} 
    & = \l[\E\l\{\frac{\partial \U(\bvartheta)}{\partial\btheta}\r\}\r]^\t = \E_1[\bar\omega(Y,\Z)\Z\{\bar\m_1(Y,\Z) - \X\}^\t] - \E_0\{\Z\bar\m_1(Y,\Z)^\t\},\\
    \E\l\{\frac{\partial \V(\bvartheta)}{\partial\btheta}\r\} 
    & = - \E_0(\Z\Z^\t).
\end{align*}
Let $[\J_u,\J_v]:=\J^{-1}_{\bvartheta}$ with $\J_u\in\R^{(p+q)\times p}$ and $\J_v\in\R^{(p+q)\times q}$. Thus, $\J_{\bvartheta},\J_u,\J_v$ are independent of $\bvartheta$. We then introduce three sets of assumptions as follows. 

\begin{assumption}[Regularity conditions]\label{assum1:regularity} 
Assume that $\bvartheta_0$ is an interior point of a compact parameter space $\Theta$.
Conditional on the discrete components of $\Z$, the continuous components of $(Y,\Z^\t)^\t$, excluding the intercept component of $\Z$, have a continuously differentiable density in both populations. There exists a constant $C_U>0$ such that $\E_s\{ \bar\omega^4(Y,\Z) + Y^4 + Y^{16} + \|\Z\|_2^4 + \|\Z\|_2^{16} +\|\X\|_2^8\} < C_U$ for $s\in\{0,1\}$. 
Moreover, $\bar\m_1(Y,\Z)$ and $\bar\m_2(Y,\Z)$ are square-integrable under both populations, and all products involving
$\bar\omega(Y,\Z)$, $\bar\m_1(Y,\Z)$, and $\bar\m_2(Y,\Z)$
that enter the source- and target-sample estimating functions
have finite second moments.
Assume that $\mu(\x\mid y,\z;\bgamma)$ is twice continuously differentiable with respect to $\bgamma$, and that differentiation under the integral sign
is valid up to the second order for $\m_\iota(y,\z;\bgamma)$,
$\iota=1,2$, in a neighbourhood $\mathcal N$ of $\bar\bgamma$.
Further, assume that, for $s\in\{0,1\}$ and $\iota=1,2$,
$
\E_s\left\{
\|\nabla_\bgamma \m_\iota(Y,\Z;\bar\bgamma)\|_2^4
+
\sup_{\bgamma\in\mathcal N}
\|\nabla_\bgamma^2 \m_\iota(Y,\Z;\bgamma)\|_2^4
\right\}<\infty.
$
The information matrix $\J_{\bvartheta_0}$ has all its eigenvalues bounded away from 0 and $\infty$. 
\end{assumption}

\begin{assumption}[Specification of the nuisance models]\label{assum2:models}
    At least one of the following two conditions holds: $(i)$ $\w(Y,\Z) = \exp(Y\eta_{y0}+\Z^\t\beeta_{z0})$ for some $\eta_{y0}$ and $\beeta_{z0}$; or $(ii)$ $\bar\m_{\iota}(Y,\Z)=\bmu_{\iota}(Y,\Z)$ for $\iota=1,2$.
\end{assumption}

\begin{assumption}[Estimation error of the nuisance models]\label{assum3:error}
The nuisance estimators admit the asymptotically linear representations
\begin{align*}
\sqrt{n}(\widehat\beeta-\bar\beeta)
&=
\frac{1}{\sqrt{n}}\sum_{i=1}^n
\boldsymbol\psi_{\eta}^{1}(Y_i,\Z_i)
+
\frac{\sqrt{n}}{N}\sum_{i=n+1}^{n+N}
\boldsymbol\psi_{\eta}^{0}(Y_i,\Z_i)
+o_p(1),\\
\sqrt{n}(\widehat\bgamma-\bar\bgamma)
&=
\frac{1}{\sqrt{n}}\sum_{i=1}^n
\boldsymbol\psi_{\gamma}(\X_i,Y_i,\Z_i)
+o_p(1),
\end{align*}
where the influence functions have mean zero and finite covariance matrices
under the corresponding populations.
\end{assumption}
 
Assumption~\ref{assum1:regularity} imposes standard smoothness and moment conditions for the asymptotic analysis of $M$-estimators
\citep[Chapter 5]{van2000asymptotic}.
Assumption~\ref{assum2:models} requires that either the density ratio model or both imputation models are correctly specified. Correct specification of the full working conditional density is sufficient but not necessary for condition~(ii). Assumption~\ref{assum3:error} requires standard asymptotically linear representations for the nuisance estimators, which hold for regular finite-dimensional $M$-estimators under conventional conditions.

The following theorem establishes consistency for the proposed estimator: 
\begin{theorem}\label{thm:main-consistent}
    Under Assumptions~\ref{assum1:regularity}-\ref{assum3:error}, it holds that 
    \[
        \|\widehat\bvartheta_{\rm DR}-\bvartheta_0\|_2 = o_p(1).
    \]
\end{theorem}

We next present the main asymptotic results. Let $\c\in\R^{p+q}$ be any unit-norm vector.
\begingroup
\allowdisplaybreaks
\begin{theorem}\label{thm:main}
    Under Assumptions~\ref{assum1:regularity}-\ref{assum3:error}, as $n,N\to\infty$ with $n/(n+N)\to\rho\in(0,1)$, it holds that 
    \begin{align}\label{eq:thm_main}
        \sqrt{n}\ \c^\t (\widehat\bvartheta_{\rm DR}-\bvartheta_0)
        & = \frac{1}{\sqrt{n}}\sumn F^1_i + \frac{\sqrt{n}}{N}\sumN F^0_i + \sqrt{n}\bxi_{\eta}^\t(\widehat\beeta-\bar\beeta) + \sqrt{n}\bxi_{\gamma}^\t(\widehat\bgamma-\bar\bgamma) + o_p(1).
    \end{align}
where $\bxi_{\eta},\bxi_{\gamma}$ are provided in Appendix~A1.5 and $F^1_i,F^0_i$ are given by
\[
\small
\begin{aligned}
    F^1_i &= \bar\omega(Y_i,\Z_i)\c^\t \J_u [\{\X_i - \bar\m_1(Y_i,\Z_i)\}(Y_i- \Z_i^\t\btheta_0) + \{\bar\m_2(Y_i,\Z_i) - \X_i\X_i^\t\}\bbeta_0]  \\
    &\quad + \bar\omega(Y_i,\Z_i)\c^\t \J_v\Z_i\{\bar\m_1(Y_i,\Z_i) - \X_i\}^\t\bbeta_0,\\
    F^0_i &= \c^\t \J_u \{ \bar\m_1(Y_i,\Z_i)(Y_i - \Z_i^\t\btheta_0) - \bar\m_2(Y_i,\Z_i)\bbeta_0\} + \c^\t \J_v\Z_i \{Y_i -\bar\m_1(Y_i,\Z_i)^\t\bbeta_0 - \Z_i^\t\btheta_0\}.
\end{aligned}
\]
Consequently,
$\sqrt{n}\,\c^\t(\widehat\bvartheta_{\rm DR}-\bvartheta_0)$
converges in distribution to a mean-zero Gaussian distribution with
variance
$
\operatorname{Var}_1\left\{
F^1
+
\bxi_\eta^\t\boldsymbol\psi_\eta^1(Y,\Z)
+
\bxi_\gamma^\t\boldsymbol\psi_\gamma(\X,Y,\Z)
\right\}
+
\rho
\operatorname{Var}_0\left\{
F^0
+
\bxi_\eta^\t\boldsymbol\psi_\eta^0(Y,\Z)
\right\}/(1-\rho).
$
\end{theorem}
\endgroup

Theorem~\ref{thm:main} establishes that the proposed estimator \(\widehat\bvartheta_{\rm DR}\) is $n^{1/2}$-consistent and asymptotically normal under the stated regularity and nuisance-model conditions. The first and second terms on the right-hand side of Eq~\eqref{eq:thm_main} represent contributions from the source and target data, respectively. When Assumption~\ref{assum2:models} (i) holds, i.e., the density ratio is correctly specified, $\bxi_{\gamma}=\zero$, which indicates that $\widehat\bgamma-\bar\bgamma$ has no impact on the asymptotic expansion of \( \widehat\bvartheta_{\rm DR}\). When Assumption~\ref{assum2:models} (ii) holds, i.e., the imputation models are correctly specified, $\bxi_{\eta}=\zero$, which indicates that $\widehat\beeta-\bar\beeta$ has no impact on the asymptotic expansion of \( \widehat\bvartheta_{\rm DR} \).

\begin{theorem}\label{thm:semiparametric_efficient}
Under Assumptions~\ref{assum1:regularity}-\ref{assum3:error}, suppose the density ratio model and both imputation models are correctly specified and $n,N\to\infty$ with $n/(n+N)\to\rho\in(0,1)$. Then $\widehat\bvartheta_{\rm DR}$ is a semiparametrically efficient estimator of $\bvartheta_0$ under the sub-population shift assumption (Eq~\eqref{assum:joint pdf}). 
The semiparametric efficiency bound for the \(\sqrt{n+N}\)-scaled estimator
$ \sqrt{n+N}\ (\widehat\bvartheta_{\rm DR}-\bvartheta_0) $ is $\J_0 \V_0\J_0^\t$, where
$
\J_0 =
\left[
\E_0\{(\X^\top,\Z^\top)^\top(\X^\top,\Z^\top)\}
\right]^{-1}
$
is the inverse of the information matrix for estimating equation~\eqref{eq:estimating_eq}, $\bphi(\x,y,\z;\bvartheta):=(\x^\t,\z^\t)^\t(y-\x^\t\bbeta-\z^\t\btheta)$,    $\bpsi(y,\z;\bvartheta):=\E\{\bphi(\X,y,\z;\bvartheta)\mid y,\z\}$, and 
{\small
\begin{align*}
    \V_0 = \frac{n+N}{N} \E_0\{\bpsi(Y,\Z;\bvartheta_0)^{\otimes 2}\} + \frac{n+N}{n}\E_1[\w^2(Y,\Z)\{\bphi(\X,Y,\Z;\bvartheta_0) - \bpsi(Y,\Z;\bvartheta_0)\}^{\otimes 2}].
\end{align*}
}
\end{theorem}

Theorem~\ref{thm:semiparametric_efficient} establishes semiparametric efficiency for completely missing covariates under sub-population shift.  
Since the theorem is stated for the \(\sqrt{n+N}\)-scaled estimator, the corresponding leading covariance of \(\widehat\bvartheta_{\rm DR}\), when both nuisance models are correctly specified, has the form
{\small
\[
    \J_0 \left[
    \frac{1}{N}\E_0\{\bpsi(Y,\Z;\bvartheta_0)^{\otimes 2}\}
    +
    \frac{1}{n}\E_1\left[
    \w^2(Y,\Z)
    \{\bphi(\X,Y,\Z;\bvartheta_0)-\bpsi(Y,\Z;\bvartheta_0)\}^{\otimes 2}
    \right]
    \right]\J_0^\t .
\]
}
This expression separates the target-sample contribution from the source-sample contribution. The first term is a target expectation and depends on the observed target variables \((Y,\Z)\) through \(\bpsi(Y,\Z;\bvartheta_0)\). The second term is the residual \(\X\)-dependent component and is averaged only over the source sample. Thus, target observations can improve the \((Y,\Z)\)-based component, but they do not in general remove the source-sample contribution induced by the completely missing covariate.
The technical proofs of Theorems~\ref{thm:main-consistent}-\ref{thm:semiparametric_efficient} are provided in Appendices~A1.4-A1.6.

Appendix~A1.7 provides an illustrative calculation in a Gaussian no-shift setting with \(p=1\) and \(\Z\in\R^q\), \(\w(y,\z)\equiv 1\), and Gaussian \((X,\Z)\). This calculation makes the \((q,n,N)\) trade-off transparent:
target observations contribute through the \((Y,\Z)\)-based component, while the residual \(\X\)-dependent component is estimable only from the source sample. It also illustrates the distinction between the larger sub-population shift model used in Theorem~\ref{thm:semiparametric_efficient} and the smaller common-population model \(p_0(y,\z)=p_1(y,\z)\), under which the \((Y,\Z)\)-based component may be estimable using all \(n+N\) observations.

\section{Simulation studies}\label{sec:simulation}
In this section, we evaluate the finite-sample performance of the proposed estimators. 
\subsection{Simulation setting}\label{subsec:sim_setting}
We fix the sum of the sample sizes of the source data ($n$) and the target data ($N$) such that $n+N=2{,}000$, and our generating mechanisms of $S_i$ ensure that the sample size ratio of the two populations $N/n$ remains within the range $(0.2,0.5)$. We consider $p=1$ and $q=2$, with $\Z=(1,Z)^\t$ and
$\btheta:=(\theta_{\rm int},\theta)^\t$. We consider univariate $X$ and $Z$ with the following multiple data generation configurations for simplicity. Multivariate simulations are reported in Appendix~A4. We first generate $(Y_i,Z_i)$ and $X_i\mid Y_i,Z_i$ for both populations. Specifically, we generate $Y_i\sim N(0,1^2)$ and $Z_i\sim N(0,2^2)$ for $i=1,\ldots,n+N$. We consider two models for $X_i\mid Y_i,Z_i$:
\[
\small
\begin{aligned}
    \mathbf{M}_{\mathrm{cor}}: \quad & X_i = -1 + Y_i - 2Z_i + \eps_i,\\
    \mathbf{M}_{\mathrm{mis}}: \quad & X_i = -1 + Y_i - 4Z_i + 0.5Y_iZ_i + \eps_i,
\end{aligned}
\]
where the noise term $\eps_i$ follows a normal distribution with mean zero and variance $\s_{\eps}^2=0.2^2$. The imputation models $m_1(y,\z)=(y,1,z)^\t\bgamma$ and $m_2(y,\z) = \s_{\eps}^2 + \{(y,1,z)^\t\bgamma\}^2$ are correctly specified under $\mathbf{M}_{\mathrm{cor}}$ but misspecified under $\mathbf{M}_{\mathrm{mis}}$. We consider two models to generate a membership variable $S_i$:
\[
\small
\begin{aligned}
    \mathbf{W}_{\mathrm{cor}}: \quad & \operatorname{logit}\{\Pr(S_i=1\mid Y_i,Z_i)\} = 1 - 0.6Y_i - 0.5Z_i,\\
    \mathbf{W}_{\mathrm{mis}}: \quad & \operatorname{logit}\{\Pr(S_i=1\mid Y_i,Z_i)\} = 2.2 - 0.6Y_i - 0.5Z_i - Y_iZ_i,
\end{aligned}
\]
where $\operatorname{logit}(a)=\log\{a/(1-a)\}$ for given $a\in(0,1)$. We assign the $i$th observation to the source population when $S_i=1$ and to the target population when $S_i=0$. Then, after splitting the observations into source and target groups, we index the source observations first so that \(S_i=I(1\le i\le n)\) for simplicity. We then remove $X_i$ from the target sample to mimic the completely missing covariate setting. The density ratio model $\omega(y,\z) = \exp\{(y,1,z)^\t\beeta\}$ is correctly specified under $\mathbf{W}_{\mathrm{cor}}$ but misspecified under $\mathbf{W}_{\mathrm{mis}}$. Finally, we have three configurations: (I) $\mathbf{M}_{\mathrm{cor}}$ and $\mathbf{W}_{\mathrm{cor}}$, (II) $\mathbf{M}_{\mathrm{mis}}$ and $\mathbf{W}_{\mathrm{cor}}$, and (III) $\mathbf{M}_{\mathrm{cor}}$ and $\mathbf{W}_{\mathrm{mis}}$. Our data generation and model specification have a similar spirit to \citet{cai2025semi}. 
Appendix~A4 additionally reports Configuration~(IV), in which both nuisance models are misspecified, to illustrate the boundary of double robustness. Our goal is to estimate $\beta$ and $\btheta$, the coefficients of $X$ and $\Z$ in the target-population working model, respectively.

We also compare the proposed doubly robust estimator with the preliminary IW and IMP estimators as two benchmark estimators. To facilitate interpretation, we do not show results for the intercept $\theta_{\text{int}}$. As with previous settings, we also conduct the above simulations using centred data of $(Y,X,Z)$, ensuring that the intercept asymptotically approaches zero when the imputation models are correctly specified. The results are consistent with those obtained using non-centred data. Detailed results are shown in Tables~A1-A2 in Appendix~A4. For completeness, we also compare with pooled implementations of methods from the missing-data literature \citep{lipsitz1999weighted,han2014multiply,kluger2025prediction}. The implementations considered here pool the source and target observations without accounting for source-target heterogeneity. Consequently, they are not generally aligned with the target-population estimand and can perform poorly in our simulations. Detailed results are reported in Tables~A7-A8 in Appendix~A4.

For each setting, we generate 500 bootstrap samples to estimate the variance of the three estimators above and 500 simulation replications to summarise the average performance measures. For the given estimators $\widehat\beta$ and $\widehat\theta$, which correspond to the coefficients of $X$ and $Z$ respectively, we report the empirical average bias, root mean squared error (RMSE), standard error, and coverage rate of the nominal 95\% confidence interval in Table~\ref{tab:point-var}.

\subsection{Results}\label{subsec:simu_result}
As shown in Table~\ref{tab:point-var}, when both nuisance models are correct (Configuration (I)), the two preliminary methods (IW and IMP) and the proposed doubly robust method show similar performance in terms of bias and RMSE. When the imputation model is misspecified (Configuration (II)), IMP exhibits a larger bias and RMSE than IW and the proposed method, whereas with a misspecified density ratio model (Configuration (III)), IW shows a greater bias and RMSE than IMP and the proposed method. However, the proposed method yields nearly unbiased estimators of $\beta$ and $\theta$ in all Configurations~(I), (II), and (III), consistent with double robustness when at least one nuisance model is correctly specified. The bootstrap standard errors of the proposed estimator generally lie between those of IW and IMP. Regarding the coverage rate, IW has poor coverage rates below the nominal level of 95\% in most cases, and IMP also has unsatisfactory coverage in Configuration~(II). For the proposed method, coverage ranges from 93.1\% to 96.4\% in five of the six parameter-configuration combinations. Its lowest coverage is 86\% for $\theta$ in Configuration~(II), compared with 81.6\% for IW and 61.7\% for IMP. A sample-size investigation for Configuration~(II) is reported in Appendix~A4, showing that the bias and RMSE of the proposed estimator decrease as the total sample size increases.

\begin{table}[!ht]
\caption{Simulation results for $\beta$ and $\theta$.}\label{tab:point-var}
\centering
\resizebox{\columnwidth}{!}{
\begin{tabular}{lccccccccccccc}
\toprule
& \multicolumn{3}{c}{\textbf{Average Bias}} & \multicolumn{3}{c}{\textbf{RMSE}} & \multicolumn{3}{c}{\textbf{Standard Error}} & \multicolumn{3}{c}{\textbf{Coverage Rate}} \\
\cmidrule(lr){2-4} \cmidrule(lr){5-7} \cmidrule(lr){8-10} \cmidrule(lr){11-13}
True & IW & IMP & Proposed & IW & IMP & Proposed & IW & IMP & Proposed & IW & IMP & Proposed \\
\midrule
\multicolumn{13}{c}{\textbf{Configuration (I)}} \\
$\beta =0.959$ & -0.002 & -0.001 & -0.001 & 0.010 & 0.006 & 0.007 & 0.009 & 0.006 & 0.007 & 0.919 & 0.950 & 0.947 \\
$\theta=1.916$ & -0.004 & -0.001 & -0.003 & 0.024 & 0.011 & 0.015 & 0.021 & 0.012 & 0.015 & 0.898 & 0.950 & 0.943 \\
\multicolumn{13}{c}{\textbf{Configuration (II)}} \\
$\beta =0.468$ & 0.008 & 0.026 & 0.000 & 0.029 & 0.040 & 0.029 & 0.020 & 0.030 & 0.026 & 0.827 & 0.838 & 0.931 \\
$\theta=1.781$ & 0.034 & 0.204 & 0.056 & 0.121 & 0.240 & 0.145 & 0.083 & 0.127 & 0.113 & 0.816 & 0.617 & 0.860 \\
\multicolumn{13}{c}{\textbf{Configuration (III)}} \\
$\beta =0.952$ & -0.003 & -0.001 & -0.001 & 0.009 & 0.006 & 0.006 & 0.008 & 0.006 & 0.006 & 0.942 & 0.961 & 0.964 \\
$\theta=1.918$ & -0.029 & -0.001 & -0.001 & 0.034 & 0.012 & 0.012 & 0.019 & 0.012 & 0.013 & 0.637 & 0.957 & 0.960 \\
\bottomrule
\end{tabular}
}
\begin{tablenotes}
    \item \footnotesize IW, importance weighting method; IMP, imputation method; Proposed, proposed doubly robust method.
\end{tablenotes}
\end{table}

To further assess the role of the identifying condition in finite samples, we also examine how the three methods behave when the sub-population shift assumption \(p_0(x\mid y,z)=p_1(x\mid y,z)\) in Eq~\eqref{assum:joint pdf} is gradually violated. In this analysis, the density ratio model for \((Y,Z)\) remains correctly specified, while the conditional distribution of \(X\mid Y,Z\) differs between the source and target populations through a perturbation parameter \(\delta\). Appendix~A4.6 gives the data-generating details and the full results. As \(\delta\) increases, Tables~A9-A10 show increasing bias and RMSE for all methods, together with deteriorating coverage. These results indicate that violations of the identifying condition induce structural bias that the correct specification of the density ratio model for \((Y,Z)\) alone cannot remove; thus, this analysis is best interpreted as a diagnostic of the identifying condition, rather than as a robustness guarantee beyond it.

\section{Case study: Experiments using UK Biobank data}\label{sec:data analysis}
\subsection{Data introduction}
In this section, we assess the proposed method using UK Biobank data (\href{https://www.ukbiobank.ac.uk/}{www.ukbiobank.ac.uk}), a large-scale biomedical resource with rich and diverse covariate information. To mimic a setting where certain covariates are entirely missing in the target population but available in the source population, we divide the data into target and source groups and explicitly examine the covariate shift between them. A key covariate is then deliberately omitted from the target population. We compare our method against two baseline approaches, IW and IMP, that were also considered in the simulation studies. This artificial masking design provides fully observed target-data benchmark estimates of the coefficients defined by Eq~\eqref{eq:estimating_eq}.

Our outcome of interest, $Y$, is body mass index (BMI), a key risk factor for numerous health conditions, including type 2 diabetes, hypertension, cardiovascular disease, and certain cancers. We analyse data from 7{,}919 White British participants under two scenarios with artificially introduced missingness. In Case I, the source population comprises 4{,}928 individuals under age 65, and the target population includes 2{,}991 individuals aged 65 and older. Covariates $\Z$ include total energy intake and sex. In Case II, the source population consists of 1{,}468 individuals with obesity (BMI $\geq$ 30), and the target population includes 6{,}451 without obesity, with $\Z$ including total energy, sex, and age, which are commonly used in BMI-related studies \citep{bray1998dietary,arem2013healthy}. 
By this construction, the supports of $Y$ in the two constructed groups are disjoint, so the formal overlap condition does not hold. We therefore include Case II as an extrapolation illustration. In both cases, we use the polygenic risk score for BMI as $X$, which is frequently unavailable in the target population due to data-collection priorities, protocol constraints, or cost. We standardise total energy and age and centre $X$ and $Y$ before fitting the models. As partial diagnostic checks for the sub-population shift assumption in Eq~\eqref{assum:joint pdf}, we regress $X$ on $Y$ and $\Z$ in the pooled sample and compare the resulting residual distributions across the source and target groups. Kolmogorov–Smirnov tests provide no evidence that the source and target residual distributions differ in either Case I ($p$-value = 0.751) or Case II ($p$-value = 0.650). Table~A13 in Appendix~A5 additionally reports separate regressions of $X$ on $Y$ and $\Z$ in the two constructed groups as a descriptive comparison of their fitted conditional mean models.

\subsection{Benchmark results with $X$ observed}\label{subsec:benchmark}
We begin by presenting the linear regression results for the two populations separately, using the observed $X$ in the target data (Table~\ref{tab:summary}). Comprehensive results are available in Appendix~A5, detailed in Table~A11. These estimates serve as benchmarks for assessing the performance of the proposed estimators.

As shown in Table~\ref{tab:summary}, there are notable differences in the estimated covariate coefficients and $p$-values between the two populations in both cases. In Case I, the coefficient of $X$ for the target population (age $\geq$ 65) is smaller than that of the source population (age $<$ 65), corresponding to a smaller estimated association between $X$ and BMI among older individuals. The estimated coefficients for energy and sex also differ between the two populations. 

In Case II, the coefficient of $X$ for the target population (BMI $<$ 30) is also smaller than that of the source population (BMI $\geq$ 30), corresponding to a smaller estimated association between $X$ and BMI among non-obese individuals than among individuals with obesity. The coefficients for sex and age have opposite signs across the two groups, further illustrating differences in their fitted regression relationships. The energy coefficient differs from zero at the 5\% level in the obese population but not in the non-obese population.

These coefficient differences illustrate source-target heterogeneity and motivate methods that estimate target-population regression parameters without directly transferring source-population coefficients.

\begin{table}[!ht]
\caption{Benchmark results with $X$ observed in both populations.}\label{tab:summary}
\begin{tabular*}{\columnwidth}{@{\extracolsep\fill}ccrrccr@{\extracolsep\fill}}
\toprule
    & Covariate & Estimate & $p$-value & Covariate & Estimate & $p$-value\\[3pt]
\midrule
   Case I & \multicolumn{3}{@{}c@{}}{\bfseries \normalsize Source (age $<$ 65)} & \multicolumn{3}{@{}c@{}}{\bfseries \normalsize Target (age $\ge$ 65)}\\
    & $X$    & 1.202 & 3.3e-74 & $X$    & 1.016 & 1.3e-37 \\ 
    & energy & 0.134 & 0.036   & energy & 0.156 & 0.050 \\ 
    & sex    & 1.158 & 1.5e-18 & sex    & 0.813 & 1.4e-07 \\
\midrule
  Case II & \multicolumn{3}{@{}c@{}}{\bfseries \normalsize Source (BMI $\ge$ 30)} & \multicolumn{3}{@{}c@{}}{\bfseries \normalsize Target (BMI $<$ 30)}\\
    & $X$    & 0.755  & 9.6e-13 & $X$   & 0.446 & 6.0e-40 \\
    & energy & 0.220  & 0.017   & energy& 0.027 & 0.424   \\
    & sex    & -0.844 & 2.2e-05 & sex   & 1.154 & 6.3e-66 \\
    & age    & -0.050 & 1.7e-04 & age   & 0.017 & 9.5e-05 \\
\bottomrule
\end{tabular*}
\begin{tablenotes}
\item \footnotesize Estimate, point estimator; the $p$-values are for two-sided tests of the null hypothesis that the corresponding regression coefficient equals zero.
\end{tablenotes}
\end{table}

\subsection{Comparison of estimation and inference across methods}
When $X$ is completely missing in the target data, we compare the proposed method to the preliminary IW and IMP methods. We consider the linear imputation model $X \sim Y + energy + sex$ for Case I and $X \sim Y + energy + sex + age$ for Case II, and denote the IMP and proposed methods as ``IMP-linear'' and ``Proposed-linear'', respectively. Tables~\ref{tab:dataresults}-\ref{tab:dataresults-robust} summarise the differences between the point estimates and the fully observed target-data benchmark estimates from Section~\ref{subsec:benchmark} (denoted as ``Difference''), together with standard errors based on bootstrap variance estimation ($B=500$), $p$-values for testing the null hypothesis that the covariate coefficient equals zero, and nominal 95\% confidence intervals. The tables report nonintercept coefficients.

As shown in Table~\ref{tab:dataresults}, the proposed method yields point estimates close to the fully observed target-data benchmark estimates in both cases. The preliminary IMP method with the linear imputation model yields similarly close point estimates. In contrast, the preliminary IW estimates differ substantially from the benchmark estimates for most covariates in both cases. For the energy coefficient in Case II, the IW analysis yields $p=0.025$, whereas the fully observed target-data benchmark analysis yields $p=0.424$.

\begin{table}[!ht]
\caption{Data analysis results for the target population with missing $X$.}\label{tab:dataresults}
\begin{tabular*}{\columnwidth}{@{\extracolsep\fill}ccccccr@{\extracolsep\fill}}
\toprule
   & Method & Covariate & Difference & SE & 95\% CI & $p$-value\\[3pt]
\midrule
  Case I
    & IW  
      & $X$    & 0.187 & 0.069  & (1.068, 1.337) & 1.2e-68 \\ 
    & & energy & -0.021 & 0.068 & (0.002, 0.267) & 0.047 \\ 
    & & sex    & 0.345 & 0.134  & (0.896, 1.421) & 5.1e-18 \\[2pt] 
    & IMP-linear 
      & $X$    & 0.002 & 0.060  & (0.901, 1.135) & 7.0e-65 \\ 
    & & energy & -0.027 & 0.077 & (-0.022, 0.279)& 0.093 \\  
    & & sex    & 0.025 & 0.152  & (0.541, 1.135) & 3.2e-08 \\[2pt] 
    & Proposed-linear
      & $X$    & 0.002 & 0.059  &(0.903, 1.133) & 3.3e-67 \\ 
    & & energy & -0.027 & 0.077 &(-0.022, 0.279)& 0.093 \\ 
    & & sex    & 0.025 & 0.152  &(0.541, 1.135) & 3.2e-08 \\
  \midrule
  Case II
    & IW  
      & $X$    & 0.309 & 0.114  & (0.532, 0.978)   & 3.0e-11 \\ 
    & & energy & 0.193 & 0.098  & (0.028, 0.412)   & 0.025 \\ 
    & & sex    & -1.997 & 0.196 & (-1.228, -0.459) & 1.7e-05 \\ 
    & & age    & -0.068 & 0.015 & (-0.080, -0.021) & 0.001 \\[2pt]
    & IMP-linear  
      & $X$    & -0.074 & 0.051 & (0.271, 0.472) & 3.7e-13 \\ 
    & & energy & -0.010 & 0.032 & (-0.046, 0.079)& 0.598 \\  
    & & sex    & -0.072 & 0.073 & (0.940, 1.224) & 2.7e-50 \\ 
    & & age    & -0.001 & 0.005 & (0.008, 0.026) & 3.3e-04 \\[2pt] 
    & Proposed-linear
      & $X$    & -0.073 & 0.074 & (0.228, 0.518)  & 4.6e-07 \\ 
    & & energy & -0.010 & 0.032 & (-0.046, 0.080) & 0.602 \\ 
    & & sex    & -0.072 & 0.073 & (0.938, 1.226)  & 4.8e-49 \\ 
    & & age    & -0.001 & 0.005 & (0.007, 0.026)  & 3.8e-04 \\ 
\bottomrule
\end{tabular*}
\begin{tablenotes}
    \item \footnotesize Difference, point estimate minus the fully observed target-data benchmark estimate; SE, standard error; 95\% CI, nominal 95\% confidence interval; IW, importance weighting method; IMP-linear, imputation method with the linear imputation model; Proposed-linear, proposed doubly robust method with the linear imputation model. The $p$-values are for two-sided tests of the null hypothesis that the corresponding regression coefficient equals zero.
\end{tablenotes}
\end{table}

\subsection{Sensitivity to the imputation-model specification}
To assess the sensitivity of the proposed method to the imputation specification, we repeat the analysis using quadratic and interaction specifications in both cases (Table~\ref{tab:dataresults-robust}). For the quadratic imputation model, we consider $X\sim Y + energy^2$ for Case I and $X\sim Y + energy^2 + sex$ for Case II, and denote the IMP and proposed methods as ``IMP-quadratic'' and ``Proposed-quadratic'', respectively. For the interaction imputation model, we consider $X\sim Y + energy*sex$ for Case I and $X\sim Y + energy*age + sex$ for Case II, and denote the IMP and proposed methods as ``IMP-interaction'' and ``Proposed-interaction'', respectively. Here, we do not present the results of the IW method, as it does not use imputation models, and its results are the same as in Table~\ref{tab:dataresults}.

\begin{table}[!ht]
\caption{Results under alternative imputation specifications for the target population with missing $X$ in the two cases.} \label{tab:dataresults-robust}
\begin{tabular*}{\columnwidth}{@{\extracolsep\fill}lcccccr@{\extracolsep\fill}}
\toprule
   & Method & Covariate & Difference & SE & 95\% CI & $p$-value\\[3pt] 
\midrule
  Case I
    & IMP-quadratic 
      & $X$    & -0.023 & 0.060 &(0.876, 1.110) & 5.4e-62 \\ 
    & & energy & -0.036 & 0.076 &(-0.029, 0.268) & 0.114 \\  
    & & sex    & -0.061 & 0.149 &(0.459, 1.045) & 4.8e-07 \\[2pt]  
    & Proposed-quadratic 
      & $X$    & -0.010 & 0.059 &(0.890, 1.120)& 9.7e-66 \\  
    & & energy & -0.029 & 0.076 &(-0.023, 0.276) & 0.097 \\  
    & & sex    & -0.010 & 0.150 &(0.509, 1.098)& 8.9e-08 \\[2pt]  
    & IMP-interaction 
      & $X$    & -0.026 & 0.060 &(0.873, 1.107)& 1.2e-61 \\ 
    & & energy & -0.042 & 0.075 &(-0.035, 0.261) & 0.133 \\   
    & & sex    & -0.062 & 0.149 &(0.458, 1.044)& 4.9e-07 \\[2pt] 
    & Proposed-interaction 
      & $X$    & -0.013 & 0.059 &(0.888, 1.118)& 2.3e-65 \\  
    & & energy & -0.031 & 0.076 &(-0.024, 0.274) & 0.101 \\   
    & & sex    & -0.012 & 0.150 &(0.506, 1.096)& 9.8e-08 \\
  \midrule
  Case II
    & IMP-quadratic 
      & $X$    & -0.070 & 0.050 & (0.277, 0.474)& 8.8e-14 \\   
    & & energy & -0.013 & 0.031 & (-0.048, 0.075) & 0.667 \\   
    & & sex    & -0.070 & 0.073 & (0.941, 1.226)& 3.8e-50 \\  
    & & age    & -0.002 & 0.005 & (0.006, 0.024)& 0.001 \\[2pt] 
    & Proposed-quadratic 
      & $X$    & -0.047 & 0.079 & (0.243, 0.554) & 4.8e-07 \\  
    & & energy & 0.005 & 0.051  & (-0.069, 0.133) & 0.535 \\  
    & & sex    & -0.091 & 0.077 & (0.911, 1.214) & 5.0e-43 \\  
    & & age    & 0.004 & 0.008  & (0.007, 0.036) & 0.004 \\[2pt]  
    & IMP-interaction 
      & $X$    & -0.069 & 0.051 & (0.277, 0.476) & 1.4e-13 \\ 
    & & energy & -0.015 & 0.031 & (-0.048, 0.073) & 0.687 \\  
    & & sex    & -0.070 & 0.073 & (0.941, 1.226) & 3.1e-50 \\ 
    & & age    & -0.002 & 0.005 & (0.006, 0.024) & 0.001 \\[2pt] 
    & Proposed-interaction 
      & $X$    & -0.046 & 0.081 & (0.242, 0.558) & 7.1e-07 \\ 
    & & energy & 0.005 & 0.058  & (-0.081, 0.145)& 0.579 \\ 
    & & sex    & -0.092 & 0.079 & (0.908, 1.217) & 1.6e-41 \\ 
    & & age    & 0.004 & 0.008  & (0.007, 0.037) & 0.005 \\ 
\bottomrule
\end{tabular*}
\begin{tablenotes}
    \item \footnotesize Difference, point estimate minus the fully observed target-data benchmark estimate; SE, standard error; 95\% CI, nominal 95\% confidence interval; IMP-quadratic/interaction, imputation method with the quadratic/interaction imputation model; Proposed-quadratic/interaction, proposed doubly robust method with the quadratic/interaction imputation model. The $p$-values are for two-sided tests of the null hypothesis that the corresponding regression coefficient equals zero.
\end{tablenotes}
\end{table} 

With $X$ observed in the target data, the linear imputation model has the smallest AIC in both cases and the smallest BIC in Case II, and is tied with the interaction model for the smallest BIC in Case I (Table~A12 in Appendix~A5). Taken together, these criteria favour the linear specification over the two alternatives, so the comparisons in Table~\ref{tab:dataresults-robust} assess sensitivity to alternative imputation specifications. For the coefficient of $X$, the parameter of primary interest, the proposed method is closer to the fully observed target-data benchmark estimates than IMP in both cases and under both alternative specifications. In Case I, this advantage extends to all reported coefficients. In Case II, IMP is marginally closer for sex and age, with small absolute differences.

\section{Conclusion and discussion}\label{sec:discussion}
We develop a transfer learning based approach for settings where some covariates are completely missing in the target population. The sub-population shift assumption permits source-target differences in the joint distribution of \((Y,\Z)\), but requires the conditional distribution \(p(\x\mid y,\z)\) of the completely missing covariates to be invariant. Thus, it is a substantive transportability condition. It is most plausible when the absence of \(\X\) in the target population is driven by study design rather than subject-level selection, and when the recorded \((Y,\Z)\) contain the main variables needed to make \(\X\) comparable across populations. In applications, additional caution is needed when important heterogeneity is not captured by \((Y,\Z)\), such as latent ancestry, study site, disease subtype, recruitment mechanism, or measurement environment. In such settings, conditioning on \((Y,\Z)\) alone may not align the conditional distribution of \(\X\) across populations. Similar concerns can arise when the source and target studies measure covariates at different resolutions or with different definitions, so that the recorded \(\Z\) omits or coarsens variables related to \(\X\). In these cases, \(p_0(\x\mid y,\z)=p_1(\x\mid y,\z)\) may be less plausible, and correct modelling of the density ratio for \((Y,\Z)\) does not by itself address the mismatch in \(\X\mid Y,\Z\). These considerations motivate diagnostic checks and sensitivity analyses for departures from the sub-population shift assumption.

This work focuses on the working linear model~\eqref{eq:workingmodel} in the target population. A natural extension is a working partially linear model, \(\E_0(Y\mid \X,\Z)=\X^\top\bbeta+g(\Z)\), where \(g(\cdot)\) is a smooth function. Using standard spline approximation results \citep{de1978practical,ma2016inference}, one may approximate \(g(\Z)\) by \(\boldsymbol B(\Z)^\top\btheta\), where $\boldsymbol B(\Z)\in\R^{d_B}$ is a vector of spline basis functions and $\btheta\in\R^{d_B}$. The target estimating equation then becomes
$
    \E_0[(\X^\top,\boldsymbol B(\Z)^\top)^\top\{Y-\X^\top\bbeta-\boldsymbol B(\Z)^\top\btheta\}]=\zero.
$
By replacing the linear regression design $\Z$ with $\boldsymbol B(\Z)$ in the score components of the augmented estimating equations~\eqref{eq:DR_estimating_1}-\eqref{eq:DR_estimating_2}, while continuing to condition the nuisance functions on the fully observed $(Y,\Z)$, one can construct an analogous doubly robust estimator. A full theoretical treatment, including the choice and growth of the spline basis, is left for future work.

Several other extensions are also worth pursuing. Extending the approach to generalised linear models requires additional development, because the nonlinear link couples the missing covariates with the regression parameters and breaks the separability exploited by the present estimating equations. Another extension is the setting in which \(\X\) is partially observed in the target population; the key question is how to combine information from target observations with measured \(\X\) and source-assisted information from the proposed transfer estimator under different missingness mechanisms. Finally, the same idea may be useful for studying gene-environment interactions when genetic measurements are expensive or unavailable in the target data, by borrowing genetic information from a source population.

\endgroup

\paragraph*{Supplementary material}
The supplementary material provides technical proofs, discusses potential nuisance models, introduces the double machine learning framework along with the corresponding theoretical results, and includes additional results on simulation and data analysis.

\bibliographystyle{abbrvnat}
\bibliography{ref}

\end{document}